\documentclass[aps,preprintnumbers,eqsecnum,amsmath,amssymb,showpacs,nofootinbib]{revtex4}
\usepackage{graphicx}% Include figure files
\usepackage{dcolumn}% Align table columns on decimal point
\usepackage{bm}% bold math
\usepackage{epsfig}

\begin{document} %HEPHY-PUB 899/11 (2011)
\title{\boldmath Pion elastic and $(\pi^0,\eta,\eta')\to\gamma\gamma^*$ transition form factors \\
in a broad range of momentum transfers}
\author{Irina Balakireva$^{1}$, Wolfgang Lucha$^{2}$, and Dmitri Melikhov$^{1,2,3}$}
\affiliation{$^1$SINP, Moscow State University, 119991, Moscow, Russia\\
$^2$HEPHY, Austrian Academy of Sciences, Nikolsdorfergasse 18, A-1050, Vienna, Austria\\
$^3$Faculty of Physics, University of Vienna, Boltzmanngasse 5, A-1090, Vienna, Austria}
%\preprint{HEPHY-PUB 910/11, UWThPh-2011-12}
\date{\today}
\begin{abstract}
We analyze $F_\pi(Q^2)$ and $F_{P\gamma}(Q^2)$, $P=\pi,\eta,\eta'$, within the local-duality (LD) 
version of QCD sum rules, which allows one to obtain predictions for hadron form factors in a broad 
range of momentum transfers. To probe the accuracy of this approximate method, we consider, 
in parallel to QCD, a potential model: in this case, the exact form factors may be calculated from the solutions 
of the Schr\"odinger equation and confronted with the results from the quantum-mechanical LD sum rule. 
On the basis of our quantum-mechanical analysis we conclude that the LD sum rule is expected to give 
reliable predictions for $F_\pi(Q^2)$ and $F_{\pi\gamma}(Q^2)$ in the region $Q^2 \ge 5-6$ GeV$^2$. 
Moreover, the accuracy of the method improves rather fast with growing $Q^2$ in this region. 
For the pion elastic form factor, the data at small $Q^2$ indicate that the LD limit may be reached already at 
relatively low values of momentum transfers, $Q^2\approx 4-8$ GeV$^2$; we therefore conclude that 
large deviations from LD in the region $Q^2=20-50$ GeV$^2$ reported in some recent theoretical analyses seem unlikely. 
The data on the ($\eta,\eta')\to\gamma\gamma^*$ form factors meet very well the expectations 
from the LD model. Surprisingly, the {\sc BaBar} results for the $\pi^0\to\gamma\gamma^*$ form factor 
imply a violation of LD growing with $Q^2$ even at $Q^2\approx 40$ GeV$^2$, at odds with the $\eta,\eta'$ case  
and the experience from quantum mechanics.
\end{abstract}
\pacs{11.55.Hx, 12.38.Lg, 03.65.Ge, 14.40.Be}
\maketitle

\section{Introduction}
\begin{figure}[!hb]
\begin{center}
\begin{tabular}{c}
\includegraphics[width=8.5cm]{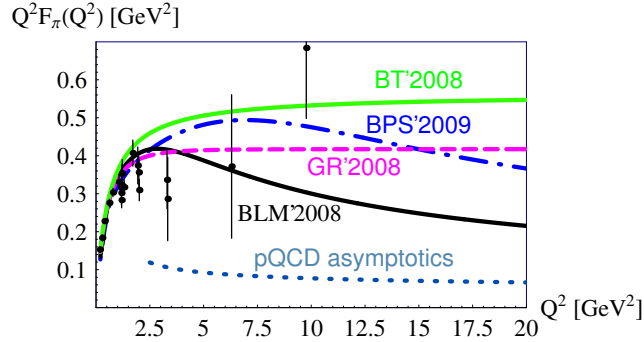}
\end{tabular}
\caption{\label{Plot:1}
Some recent theoretical predictions for the pion elastic form factor \cite{blm2008,recent} vs. the data \cite{data_piff}.}
\end{center}
\end{figure}

In spite of the long history of theoretical investigations of the pion, its properties are still not 
fully understood. For instance, no consensus on the behaviour of the pion elastic form factor
in the region  $Q^2\approx 5-50$ GeV$^2$ has been reached up to now (see Fig.~\ref{Plot:1}); the recent {\sc BaBar} results 
on the $\pi\to\gamma\gamma^*$ transition form factor \cite{babar} imply a strong violation of pQCD factorization 
in the region of $Q^2$ up to 40 GeV$^2$. 

At asymptotically large momentum transfers, $Q^2\to\infty$, the form factors satisfy QCD factorization theorems 
\cite{pqcd,bl1980} 
\begin{eqnarray} 
\label{factorization}
F_{\pi}(Q^2) \to  8\pi\alpha_s(Q^2)f_\pi^2/Q^2,\qquad F_{\pi\gamma}(Q^2)\to  \sqrt{2} f_\pi/Q^2, \qquad  f_\pi=130 \mbox{ MeV}.   
\end{eqnarray} 
However, the behaviour of the form factors at practically accessible large momentum transfers is still a subject of
lively discussions. 

Here we study $F_\pi(Q^2)$ and $F_{P\gamma}(Q^2)$ by making use of the local-duality (LD) version of QCD sum rules 
\cite{ld}. An attractive feature of this approach is the possibility to obtain predictions for hadron 
form factors in a broad range of momentum transfers without knowing subtle details of their 
structure. However, because of the approximate character of predictions from LD sum rules, 
it is important to understand the expected accuracy of the form factors obtained by this method. 
Quantum-mechanical potential models provide a possibility to probe this accuracy: one calculates the exact 
form factors by making use of the solutions of the Schr\"odinger equation and confronts these results with the 
application of LD sum rules in quantum mechanics. 

This paper is organized as follows: 
In the next section, we briefly recall some details of LD sum rules in QCD and of the LD model for form factors. 
Section 3 studies the accuracy of the LD model for elastic and transition form factors in a quantum-mechanical potential model. 
The pion elastic form factor is discussed in Section 4 and the $P\to\gamma\gamma^*$ transition from factors 
are considered in Section 5. Section 6 gives our conclusions.

%%%%%%%%%%%%%%%%%%%%%%%%%%%%%%%%%%%%%%%%%%%%%%%%%%%%%%%%%%%%%%%%%%%%%%%%%%%%%%%%%%%%%%%%%%%%%%%
\section{Local-duality model for form factors in QCD}
A local-duality sum rule \cite{ld} is a dispersive three-point sum rule at $\tau=0$ 
(i.e., infinitely large Borel mass parameter). 
In this case all power corrections vanish and the details of the non-perturbative dynamics are hidden 
in one quantity -- the effective threshold $s_{\rm eff}(Q^2)$. 

The basic object for the calculation of the pion elastic form factor 
is the vacuum $\langle AVA\rangle$ correlator, whereas for the transition form factor it is the $\langle AVV\rangle$ 
correlator, $A$ being the axial and $V$ the vector current. Implementing duality in the standard way, the sum rules  
relate the pion form factors to the low-energy region of Feynman diagrams of perturbation theory: 
\begin{eqnarray} 
\label{Fld1}
F^{\rm LD}_{\pi}(Q^2)&=& \frac{1}{f_\pi^2}\int\limits_0^{s_{\rm eff}(Q^2)}ds_1\int\limits_0^{s_{\rm eff}(Q^2)}ds_2 
\;\Delta_{\rm pert}(s_1, s_2, Q^2),\\
\label{Fld2}
F^{\rm LD}_{\pi\gamma}(Q^2)&=&\frac{1}{f_\pi}\int\limits_0^{\bar s_{\rm eff}(Q^2)}ds\;\sigma_{\rm pert}(s, Q^2).    
\end{eqnarray} 
Here $\Delta_{\rm pert}(s_1,s_2,Q^2)$ is the double spectral density 
of the $\langle AVA\rangle$ 3-point function;  
$\sigma_{\rm pert}(s,Q^2)\equiv \sigma_{\rm pert}(s,q_1^2=0,q_2^2=-Q^2)$ is the single spectral density of the 
$\langle AVV \rangle$ 3-point function. These quantities are calculated as power series in $\alpha_s$: 
\begin{eqnarray} 
\label{densities}
\Delta_{\rm pert}(s_1, s_2, Q^2)&=&\Delta^{(0)}_{\rm pert}(s_1, s_2, Q^2)+\alpha_s\Delta^{(1)}_{\rm pert}(s_1, s_2, Q^2)+O(\alpha_s^2), 
\\
\sigma_{\rm pert}(s, Q^2)&=&\sigma^{(0)}_{\rm pert}(s, Q^2)+O(\alpha_s^2).    
\end{eqnarray}
The one-loop spectral densities $\Delta^{(0)}_{\rm pert}$ and $\sigma^{(0)}_{\rm pert}$ are well-known \cite{ld,anisovich,m,ter,ms}. 
The two-loop contribution $\Delta^{(1)}_{\rm pert}$ has been calculated in \cite{bo}; 
the two-loop $O(\alpha_s)$ correction to $\sigma_{\rm pert}$ was found to be zero \cite{2loop}. 
Higher-order radiative corrections to $\sigma_{\rm pert}$ are unknown, but may not be identically zero \cite{lm2011}.  

As soon as one knows the effective thresholds $s_{\rm eff}(Q^2)$ and $\bar s_{\rm eff}(Q^2)$,  
Eqs.~(\ref{Fld1}) and (\ref{Fld2}) provide the form factors. 
However, finding a reliable criterion for fixing the thresholds is a 
very subtle problem investigated in great detail in \cite{lms1}.\footnote{It might be useful to recall 
that the ratio of the $O(1)$ and $O(\alpha_s)$ contributions to the pion elastic form factor is not 
sensitive to the details of the effective threshold and may be predicted 
with high accuracy \cite{blm2008}. In particular, at $Q^2=20$ GeV$^2$ the $O(1)$ and $O(\alpha_s)$ 
terms give approximately equal contributions to the pion form factor.} 

Due to properties of the spectral functions (see, e.g., \cite{blm2011} and references therein for details), 
the LD form factors (\ref{Fld1}) and (\ref{Fld2}) obey the factorization theorems (\ref{factorization}) 
if the effective thresholds satisfy the following relations:  
\begin{eqnarray}
\label{ass}
s_{\rm eff}(Q^2\to\infty)=\bar s_{\rm eff}(Q^2\to\infty)={4\pi^2f_\pi^2}. 
\end{eqnarray}
For finite $Q^2$, however, the effective thresholds $s_{\rm eff}(Q^2)$ and $\bar s_{\rm eff}(Q^2)$ depend on $Q^2$ 
and differ from each other \cite{lms2}. The ``conventional LD model'' arises if one assumes (\ref{ass}) 
for all ``not too small'' values of $Q^2$ \cite{ld}: 
\begin{eqnarray}
\label{sld}
s_{\rm eff}(Q^2)=\bar s_{\rm eff}(Q^2)={4\pi^2f_\pi^2}. 
\end{eqnarray}
Obviously, the LD model (\ref{sld}) for the effective continuum thresholds is an approximation which 
does not take into account details of the confinement dynamics. 
The only property of theory relevant for this model is factorization of hard form factors.

%%%%%%%%%%%%%%%%%%%%%%%%%%%%%%%%%%%%%%%%%%%%%%%%%%%%%%%%%%%%%%%%%%%%%%%%%%%%%%%%%%%%%%%%%%%%%%
%\newpage

\section{Exact vs. LD form factors in quantum-mechanical potential models}
To probe the accuracy of the LD model, we now consider a quantum-mechanical example: 
the corresponding form factors may be calculated using the solution of the Schr\"odinger equation 
and confronted with the results of the quantum-mechanical LD model, which is constructed 
precisely the same way as in QCD. For the elastic form factor, it is mandatory 
to consider a potential involving both the Coulomb and the confining parts; 
for the analysis of the transition form factor one may start with a purely confining potential. 

The basic object for quantum-mechanical LD sum rules is the analogue of the three-point correlator of field theory \cite{lms2}
\begin{eqnarray}
\label{3pointQM}
\Gamma^{\rm NR}(E,E',Q)=\langle r'=0|\frac{1}{H-E'} J(\bm{q})\frac{1}{H-E} |r=0 \rangle, 
\qquad Q\equiv |\bm{q}|. 
\end{eqnarray}
$H$ is the Hamiltonian of the model; the current operator $J(\bm{q})$ is determined by its 
kernel $\langle\bm{r}'|J(\bm{q})|\bm{r}\rangle=
\exp(i\bm{q}\cdot\bm{r})\,\delta^{(3)}(\bm{r}-\bm{r}')$. 
We do not take the spin of the current into account, therefore the basic quantum-mechanical Green function 
is the same for both types of form factors discussed above.

%%%%%%%%%%%%%%%%%%%%%%%%%%%%%%%%%%%%%%%%%
\subsection{Elastic form factor}
\begin{figure}[b!]
\begin{center}
\begin{tabular}{cc}
\includegraphics[width=12cm]{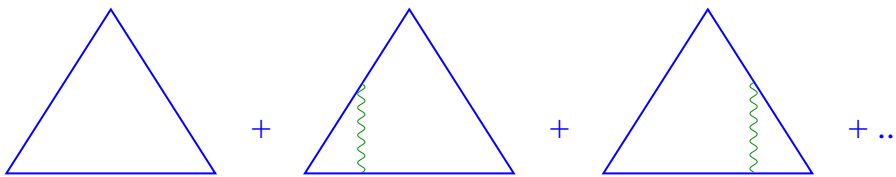}
\end{tabular}\caption{\label{Fig:2}
Feynman diagrams for the perturbative contributions to three-point 
functions in nonrelativistic~field theory. Wavy lines indicate the Coulomb potential.}
\end{center}
\end{figure}

The elastic form factor of the ground state is given in terms of its wave function~$\Psi$~by
\begin{eqnarray}
\label{QM_ff}
F_{\rm el}(Q)=\langle\Psi|J(\bm{q})|\Psi\rangle =
\int{\rm d}^3r\,\exp({\rm i}\bm{q}\cdot\bm{r})\,|\Psi(\bm{r})|^2=\int{\rm d}^3k\Psi(\bm{k})
\,\Psi(\bm{k}+\bm{q}),\qquad Q\equiv|\bm{q}|. 
\end{eqnarray}
Here, $\Psi$ is the ground state of the Hamiltonian
\begin{equation}
\label{QM_H}H=\frac{\bm{k}^2}{2m}-\frac{\alpha}{r}+V_{\rm conf}(r),\qquad r\equiv|\bm{r}|.
\end{equation}
Because of the presence of the Coulomb interaction in the potential, the asymptotic behaviour of the form factor at 
large values of $Q$ is given by the factorization theorem \cite{brodsky}
\begin{equation}
\label{QM_factorization}
F_{\rm el}(Q)\xrightarrow[Q\to\infty]{}%F_\infty(Q)\equiv
\frac{16\pi\,\alpha\,m\,R_g}{Q^4},\qquad
R_g\equiv|\Psi(\bm{r}=\bm{0})|^2.
\end{equation}
The quantum-mechanical LD sum rule for the form factor $F_{\rm el}(Q)$ is rather similar to that~in~QCD: 
The double Borel transform ($E\to T$, $E'\to T'$) of (\ref{3pointQM}) may be written in the form  
\begin{eqnarray}
\label{3pointQM_Borel}
\Gamma^{\rm NR}(T,T',Q)=
\int dk'\,\exp\left(-\frac{k'^2}{2m}T'\right)
\int dk\,\exp\left(-\frac{k^2}{2m}T\right)
\,\Delta^{\rm NR}_{\rm pert}(k,k',Q)+\Gamma^{\rm NR}_{\rm power}(T,T',Q), 
\end{eqnarray}
where $\Gamma^{\rm NR}_{\rm power}(T,T',Q)$ describes the contribution of the confining interaction and 
$\Delta^{\rm NR}_{\rm pert}(k,k',Q)$ is 
the double spectral density of Feynman diagrams of nonrelativistic perturbation theory, Fig.~\ref{Fig:2}. 

Setting $T'=T=0$ leads to the LD sum rule, in which case $\Gamma^{\rm NR}_{\rm power}$ vanishes \cite{lms1}. 
The low-energy region of perturbative diagrams---below some effective continuum threshold 
$k_{\rm eff}(Q)$---is assumed to be dual to the 
ground-state contribution, which reads $R_g\,F_{\rm el}(Q^2)$. Finally, we arrive at the following LD 
expression for the elastic form factor: 
\begin{eqnarray}
%\label{2pt_QM}
%R_g&=&\int_0^{k_{\rm eff}}{\rm d}k\,\rho^{\rm QM}_{\rm pert}(k)=\frac{k^3_{\rm eff}}{6\pi^2}
%+\alpha\,m\,\frac{k^2_{\rm eff}}{9\pi}+O(\alpha^2),\\
\label{3pt_QM}
F^{\rm LD}_{\rm el}(Q)&=&\frac{1}{R_g}\int\limits_0^{k_{\rm eff}(Q)}\,{\rm d}k\int\limits_0^{k_{\rm eff}(Q)}\,{\rm d}k'\,
\Delta^{\rm NR}_{\rm pert}(k,k',Q).
\end{eqnarray}
The rather lengthy explicit result for $\Delta^{\rm QM}_{\rm pert}(k_1,k_2,Q)$ will not be given here.

The factorization formula (\ref{QM_factorization}) is reproduced by the LD sum rule (\ref{3pt_QM}) 
if the momentum-dependent effective threshold behaves as  
\begin{equation}
\label{QM_LDa}
k_{\rm eff}(Q)\xrightarrow [Q\to\infty]{}k_{\rm LD}\equiv(6\pi^2\,R_g)^{1/3}.
\end{equation}

%%%%%%%%%%%%%%%%%%%%%%%%%%%%%%%%%%%%%%%%%%%%%%%%%%%%%%%

\subsection{Transition form factor}
The analogue of the $\pi\gamma$ transition form factor in quantum mechanics is given by
\begin{eqnarray}
\label{fnr}
F_{\rm trans}(Q, E)=\langle\Psi|J(\bm{q})\frac{1}{H-E}|r=0\rangle,
\end{eqnarray}
The case of one real and one virtual photon corresponds to $E=0$ and $Q\ne 0$. 
At large $Q$, the transition form factor $F_{\rm trans}(Q)\equiv F_{\rm trans}(Q,E=0)$ satisfies the factorization theorem 
\begin{equation}
\label{QM_factorization2}
F_{\rm trans}(Q)\xrightarrow[Q\to\infty]
{}%F_\infty(Q)\equiv%
\frac{2m\sqrt{R_g}}{Q^2}.
\end{equation}
Recall that the behaviour (\ref{QM_factorization2}) does not require the Coulomb potential in the interaction 
and---in distinction to the factorization of the elastic form factor---emerges also for a purely confining interaction. 

The LD sum rule for the form factor $F_{\rm trans}(Q)$ is constructed on the basis of the same three-point function 
(\ref{3pointQM}) and has the form 
\begin{eqnarray}
\label{fnrdual}
F^{\rm LD}_{\rm trans}(Q)=\frac{1}{\sqrt{R_g}}\int\limits_0^{\bar k_{\rm eff}(Q)}\,{\rm d}k
\int\limits_0^{\infty}{\rm d}k'\,\Delta^{\rm QM}_{\rm pert}(k,k',Q).
\end{eqnarray}
Notice that the $k'$-integration is not restricted to the low-energy region since we do not isolate the ground-state 
contribution in the initial state. 
The asymptotical behaviour (\ref{QM_factorization2}) is correctly reproduced by Eq.~(\ref{fnrdual}) for 
\begin{equation}
\label{QM_LD2}
\bar k_{\rm eff}(Q\to\infty)=k_{\rm LD}.
\end{equation}

%%%%%%%%%%%%%%%%%%%%%%%%%%%%%%%%%%%%%%%%%%%%%%%%%%%%%%%%

\subsection{Quantum-mechanical LD model}
As is obvious from (\ref{QM_LDa}) and (\ref{QM_LD2}), the effective thresholds for the elastic and for the 
transition form factors have the same limit at large $Q$: 
\begin{equation}
\label{QM_LD3}
k_{\rm eff}(Q\to\infty)=\bar k_{\rm eff}(Q\to\infty)=k_{\rm LD}.
\end{equation}
The LD model emerges when one {\it assumes} that also for intermediate $Q$ one may find a reasonable estimate 
for the form factors by setting 
\begin{equation}
\label{QM_LD}
k_{\rm eff}(Q)=\bar k_{\rm eff}(Q)=k_{\rm LD}.
\end{equation}
Similarly to QCD, the only property of the bound state which determines the form factor in the LD model is $R_g$.

%%%%%%%%%%%%%%%%%%%%%%%%%%%%%%%%%%%%%%%%%%%%%%%%%%%%%%%%
\subsection{LD vs. exact effective threshold}
Let us now calculate the exact thresholds $k_{\rm eff}(Q)$ and $\bar k_{\rm eff}(Q)$ which reproduce the exact 
form factor by the LD expression; they are obtained by solving  
the LD sum rules (\ref{3pt_QM}) and (\ref{fnrdual}) using the exact form factors on the left-hand sides of these equations. 
The deviation of the LD threshold $k_{\rm LD}$ from these exact thresholds measures the error induced by the approximation 
(\ref{QM_LD}) and characterizes the accuracy of the LD model. 

For our numerical analysis we use parameter values relevant for hadron physics: 
$m=0.175$ GeV for the reduced constituent light-quark mass and $\alpha=0.3$. 
We considered several confining potentials 
\begin{equation}
\label{QM_conf}
V_{\rm conf}(r)=\sigma_n\,(m\,r)^n, \qquad n=2,1,1/2, 
\end{equation}
and adapt the strengths $\sigma_n$ in our confining interactions such that the Schr\"odinger equation yields for each 
potential the same value of the wave function at the origin, $\Psi({r}={0})=0.078$ GeV$^{3/2},$ which holds 
for $\sigma_2=0.71$ GeV, $\sigma_1=0.96$ GeV, and $\sigma_{1/2}=1.4$~GeV. 
The ground state then has a typical hadron size $\sim$ 1 fm. 
\begin{figure}[!ht]
\begin{center}
\begin{tabular}{cc}
\includegraphics[width=8.5cm]{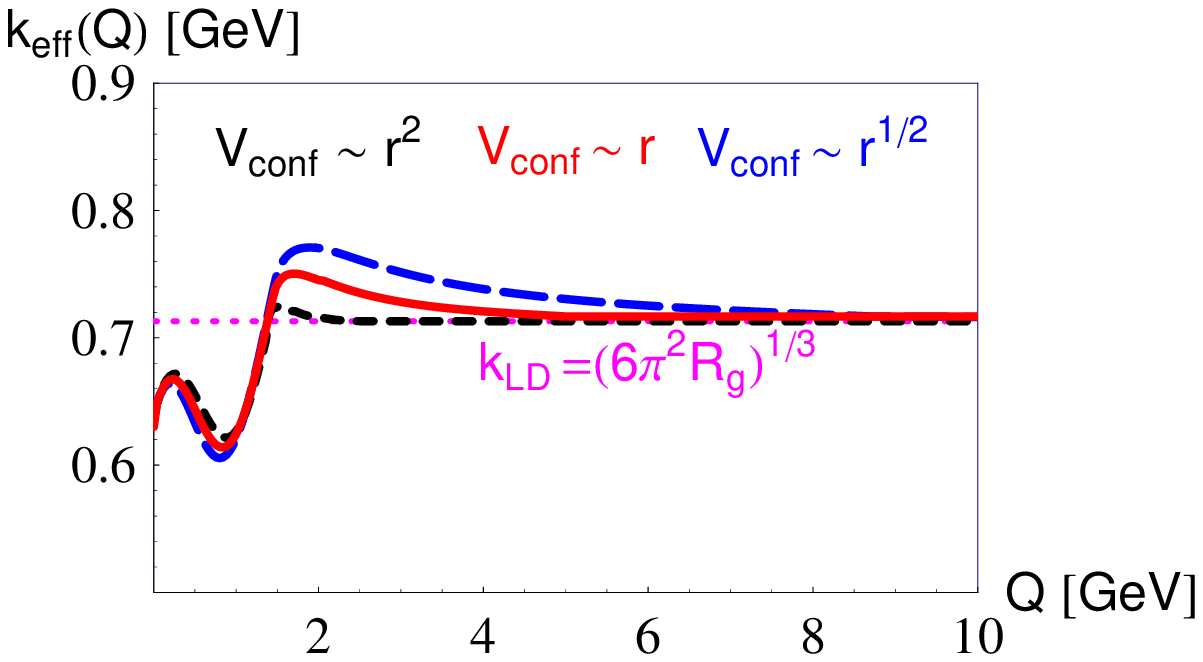}&
\includegraphics[width=8.5cm]{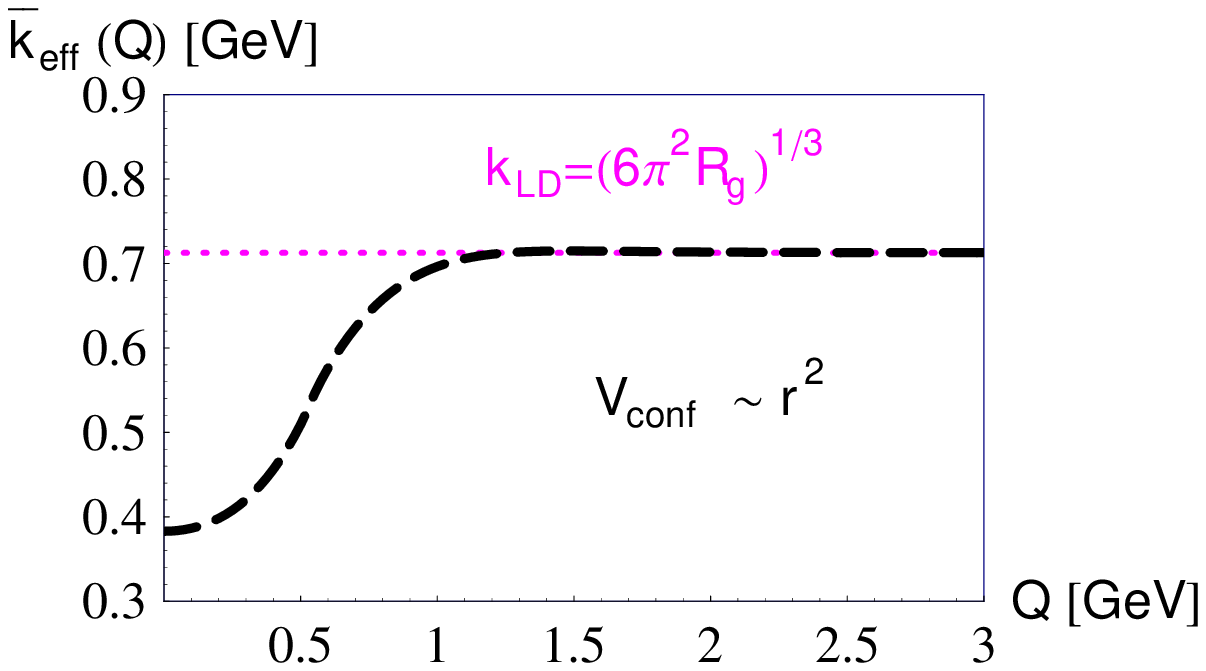}
\end{tabular}
\caption{\label{Plot:2}
Exact effective thresholds in quantum mechanics for the elastic (left) and the transition (right) form factors for 
different confining potentials. $R_{\rm g}\equiv |\Psi(r=0)|^2$.}
\end{center}
\end{figure}
Figure \ref{Plot:2} presents the exact effective thresholds. 
Independently of the details of the confining interaction, the accuracy of the LD approximation for the effective threshold 
and, respectively, the accuracy of the LD elastic form factor increases with $Q$ in the region $Q^2\ge 5-8$ GeV$^2$. 
For the transition form factor, the LD approximation works well starting with even smaller values of $Q$.

%%%%%%%%%%%%%%%%%%%%%%%%%%%%%%%%%%%%%%%%%%%%%%%%%%%%%%%%%%%%%%%%%%%%%%%%%%%%%%%%%%%%%%%%%%%%%%%%%%%%%%%%%%%%%%%%%%

%\newpage

\section{The pion elastic form factor}
\begin{figure}[!hb]
\begin{center}
\begin{tabular}{cc}
\includegraphics[width=8.5cm]{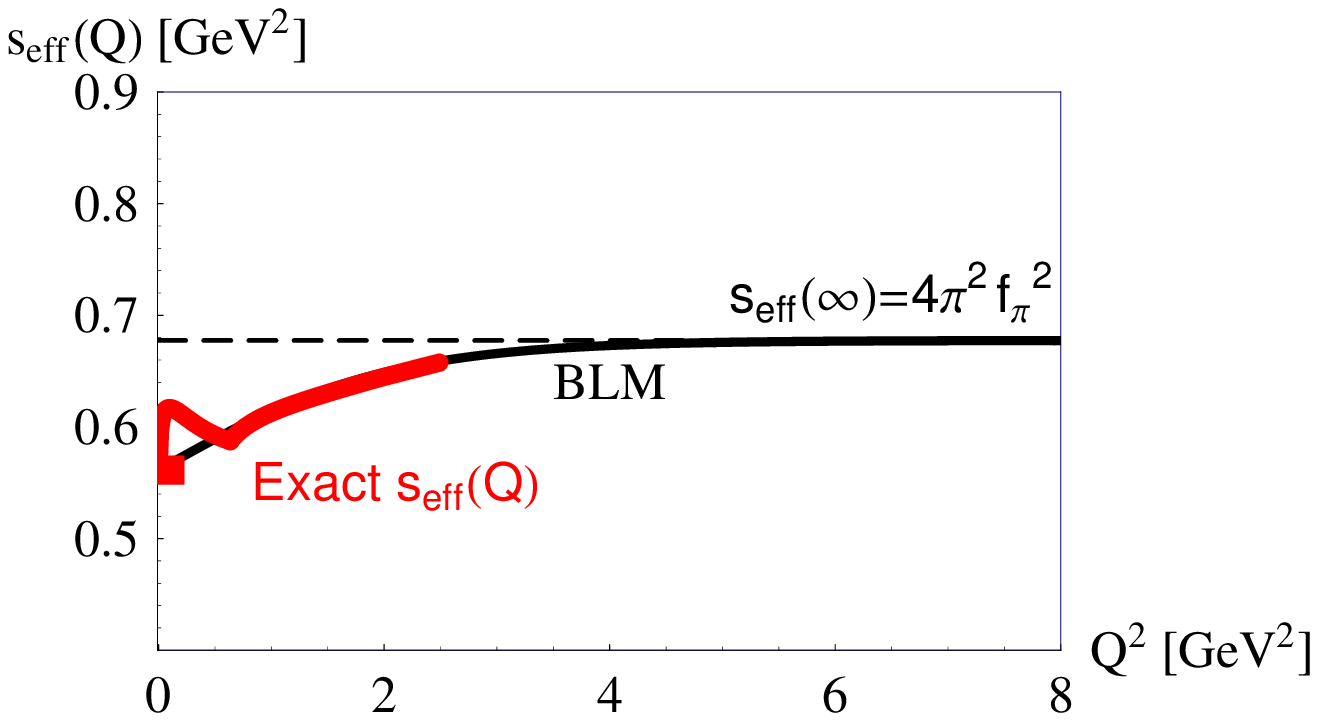}&
\includegraphics[width=8.5cm]{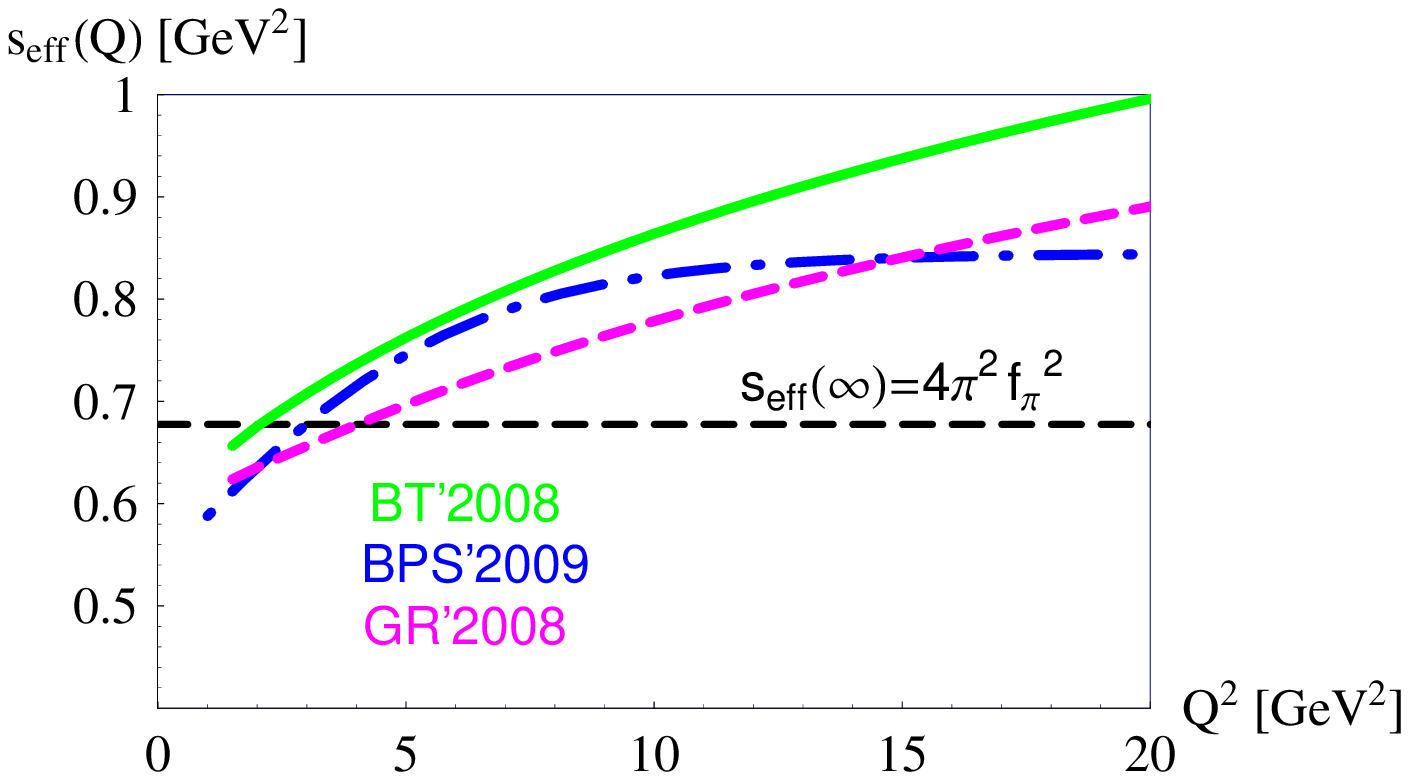}
\end{tabular}
\caption{\label{Plot:3}
Left: the ``equivalent effective threshold'' extracted from the data (red) vs. the improved LD model (BLM) 
of \cite{blm2011}. Right: equivalent thresholds for the theoretical predictions displayed in Fig.~\protect\ref{Plot:1}. }
\end{center}
\end{figure}

For a given result for the pion form factor, we define the equivalent effective threshold as the quantity which 
reproduces this result by Eq.~(\ref{Fld1}). 
Figure~\ref{Plot:3} displays the equivalent effective thresholds recalculated from the data and from the 
theoretical predictions for the elastic form factor from Fig.~\ref{Plot:1}. 
The exact effective threshold extracted from the accurate data at low $Q^2$ suggests that the LD limit may be 
reached already at relatively low values of $Q^2\approx 4-8$ GeV$^2$. However, the results in the right plot 
imply that the accuracy of the LD model still does not increase---or even decreases---with $Q^2$ even 
in the region $Q^2\simeq 20$ GeV$^2$, in conflict with both our experience from quantum mechanics and  
the hint from the data at low $Q^2$. 
We look forward to the future accurate data expected from JLab in the range up to $Q^2=8$ GeV$^2$. 

%\newpage
%%%%%%%%%%%%%%%%%%%%%%%%%%%%%%%%%%%%%%%%%%%%%%%%%%%%%%%%%%%%%%%%%%%%%%%%%%%%%%%%%%%%%%%%%%%%%%%%%%%%%%%%%%%%%%%%%%
\section{The {\boldmath $(\pi^0,\eta,\eta')\to\gamma\gamma^*$} transition form factors}
\subsection{\boldmath $(\eta,\eta')\to\gamma\gamma^*$}
Before discussing the $\pi^0$ case, let us consider the $\eta$ and $\eta'$ decays. 
Here, one has to take properly into account the $\eta-\eta'$ mixing and the presence of two---strange and 
nonstrange---LD form factors. Following \cite{anisovich,feldmann}, we describe the flavor structure 
of $\eta$ and $\eta'$ as\footnote{For comparison with the form factors obtained in a scheme based on the 
octet--singlet mixing, we refer to \cite{kot}.} 
\begin{eqnarray}
|\eta\rangle  = |\frac{\bar uu+\bar dd}{\sqrt{2}} \rangle\cos \phi-|\bar ss\rangle \sin \phi,\quad %\nonumber\\
|\eta'\rangle = |\frac{\bar uu+\bar dd}{\sqrt{2}} \rangle\sin \phi+|\bar ss\rangle \cos \phi,\quad  \phi\approx 39.3^0. 
\end{eqnarray}
The $\eta$ and $\eta'$ form factors then take the form 
\begin{eqnarray}
\label{Feta}
F_{\eta\gamma}(Q^2)=
\frac{5}{3\sqrt2}F_n(Q^2) \cos \phi-\frac{1}{3} F_s(Q^2)\sin\phi,\quad
F_{\eta'\gamma}(Q^2)=
\frac{5}{3\sqrt2}F_n(Q^2)\sin \phi+\frac{1}{3}F_s(Q^2)\cos\phi.
\nonumber\\ 
\end{eqnarray}
Here, $F_n(Q^2)$ and $F_s(Q^2)$ are the form factors describing the transition of the nonstrange 
and $\bar ss$-components, respectively. 
The LD expressions for these quantities read 
\begin{eqnarray}
F_{n\gamma}(Q^2)=\frac{1}{f_n}\int\limits_0^{s_{\rm eff}^{(n)}(Q^2)}ds\,\sigma^{(n)}_{\rm pert}(s,Q^2),\quad %\nonumber\\
F_{s\gamma}(Q^2)=\frac{1}{f_s}\int\limits_{4m_s^2}^{s_{\rm eff}^{(s)}(Q^2)}ds\,\sigma^{(s)}_{\rm pert}(s,Q^2), 
\end{eqnarray}
where $\sigma^{(n)}_{\rm pert}$ and $\sigma^{(s)}_{\rm pert}$ denote $\sigma_{\rm pert}$ with the corresponding 
quark propagating in the loop. In numerical calculations we set $m_u=m_d=0$ and $m_s=100$ MeV.  
The LD model involves two separate effective thresholds for the nonstrange and 
the strange components \cite{feldmann}:
\begin{eqnarray}
s_{\rm eff}^{(n)}&=4\pi^2f_n^2,\qquad
f_n\approx1.07f_\pi,\qquad 
s_{\rm eff}^{(s)}&=4\pi^2f_s^2,\qquad f_s\approx1.36f_\pi.
\end{eqnarray}
According to the experience from quantum mechanics, the LD model may not perform well for small values of $Q^2$, 
where the true effective threshold is smaller than the LD threshold; however, for larger $Q^2$ the LD model in 
quantum mechanics gives accurate predictions for the form factors, as illustrated by~Fig.~\ref{Plot:3}. 
\begin{figure}[!hb]
\begin{center}
\begin{tabular}{cc}
\includegraphics[width=8.5cm]{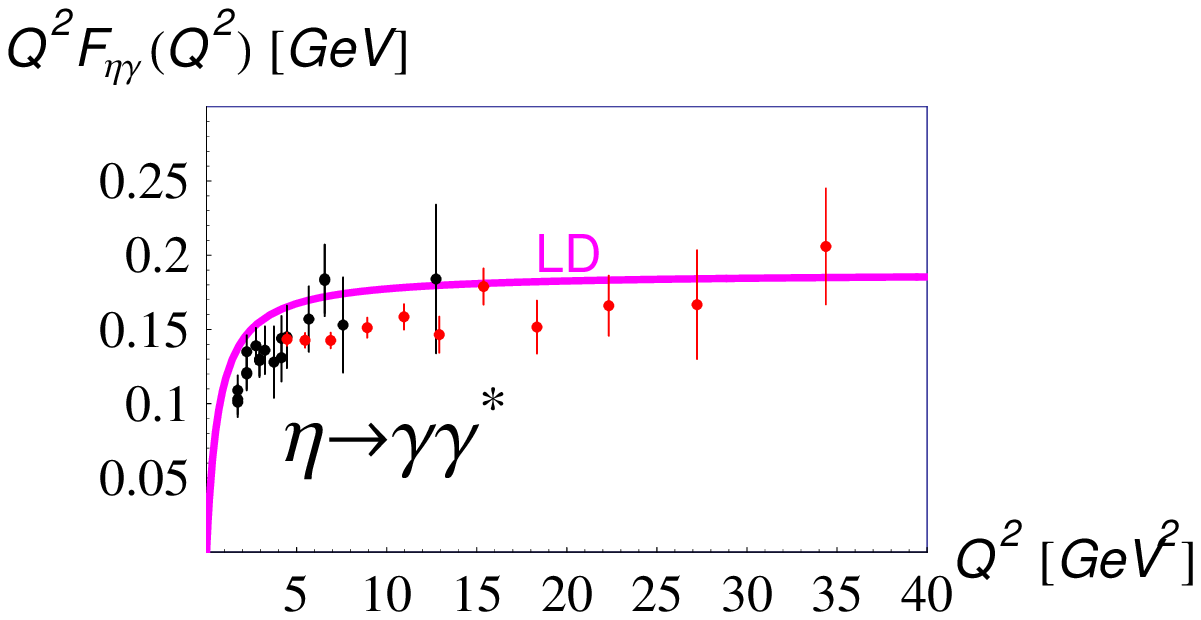} & 
\includegraphics[width=8.5cm]{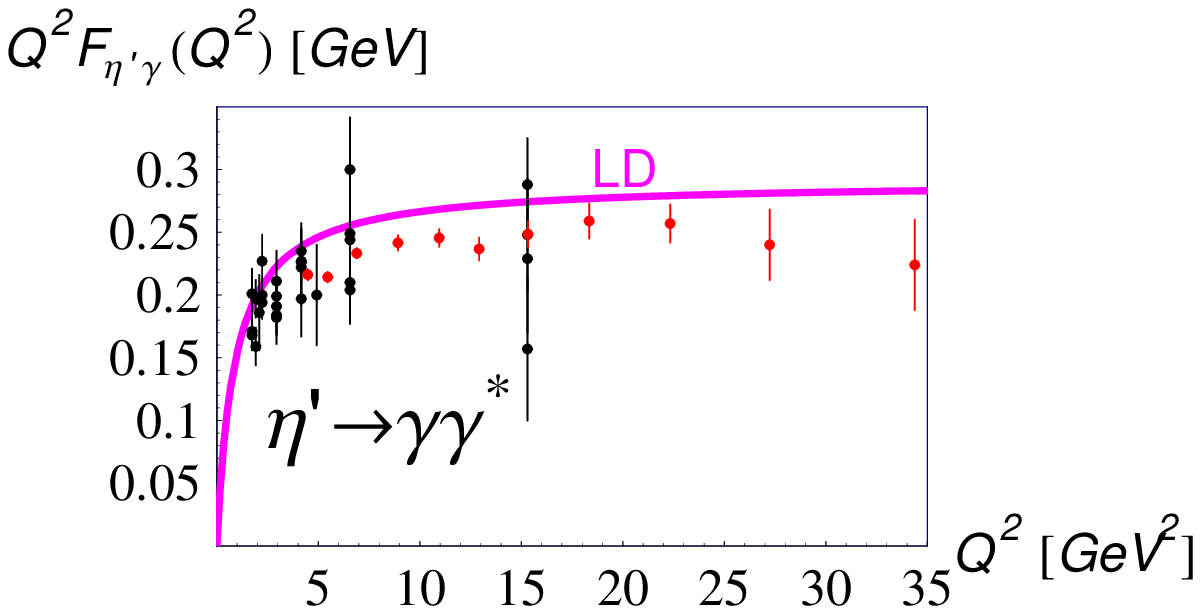}
\end{tabular}
\caption{\label{Plot:4}
LD predictions for $\eta$ and $\eta'$ vs. experimental data from \cite{cello-cleo} (black) and \cite{babar1} (red).}
\end{center}
\end{figure}
Figure~\ref{Plot:4} shows the corresponding predictions for $\eta$ and $\eta'$ mesons. One observes an overall 
agreement between the LD model and the data, meeting the expectation from quantum mechanics. 

\subsection{\boldmath $\pi^0\to\gamma\gamma^*$}
\begin{figure}[!ht]
\begin{center}
\begin{tabular}{cc}
\includegraphics[width=8.5cm]{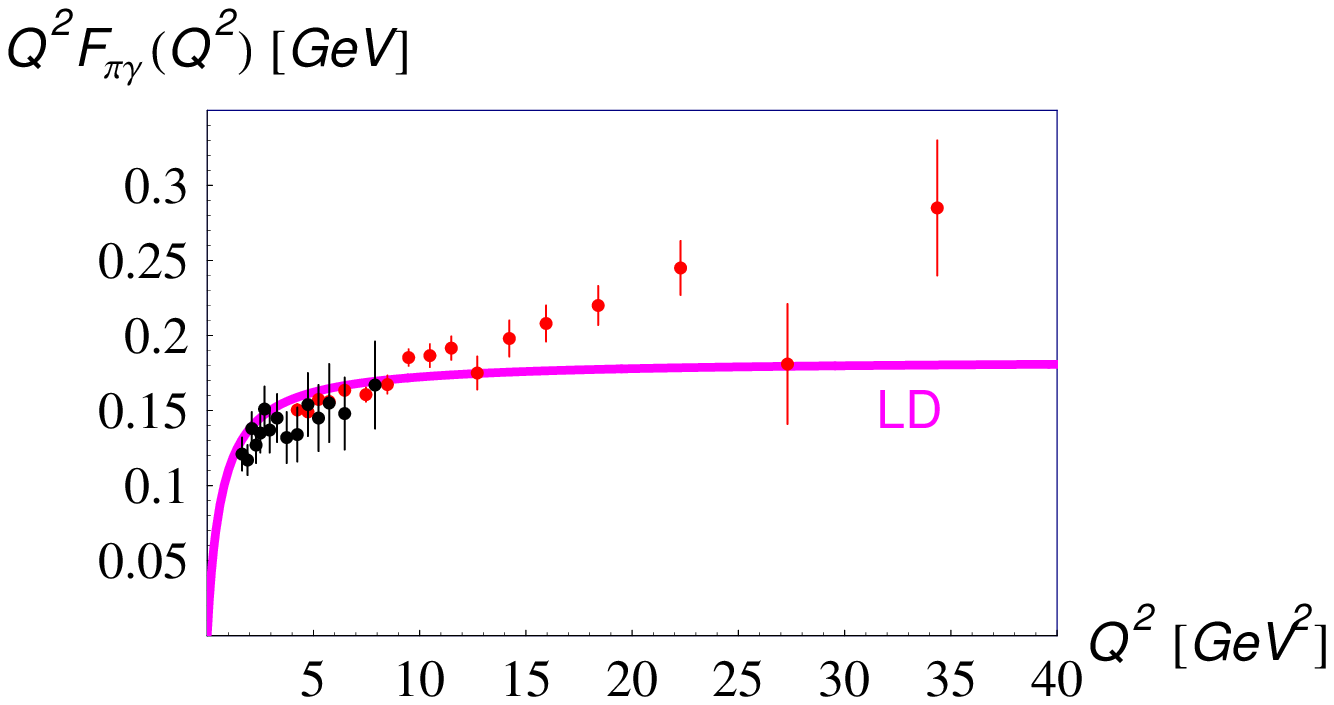}&
\includegraphics[width=8.5cm]{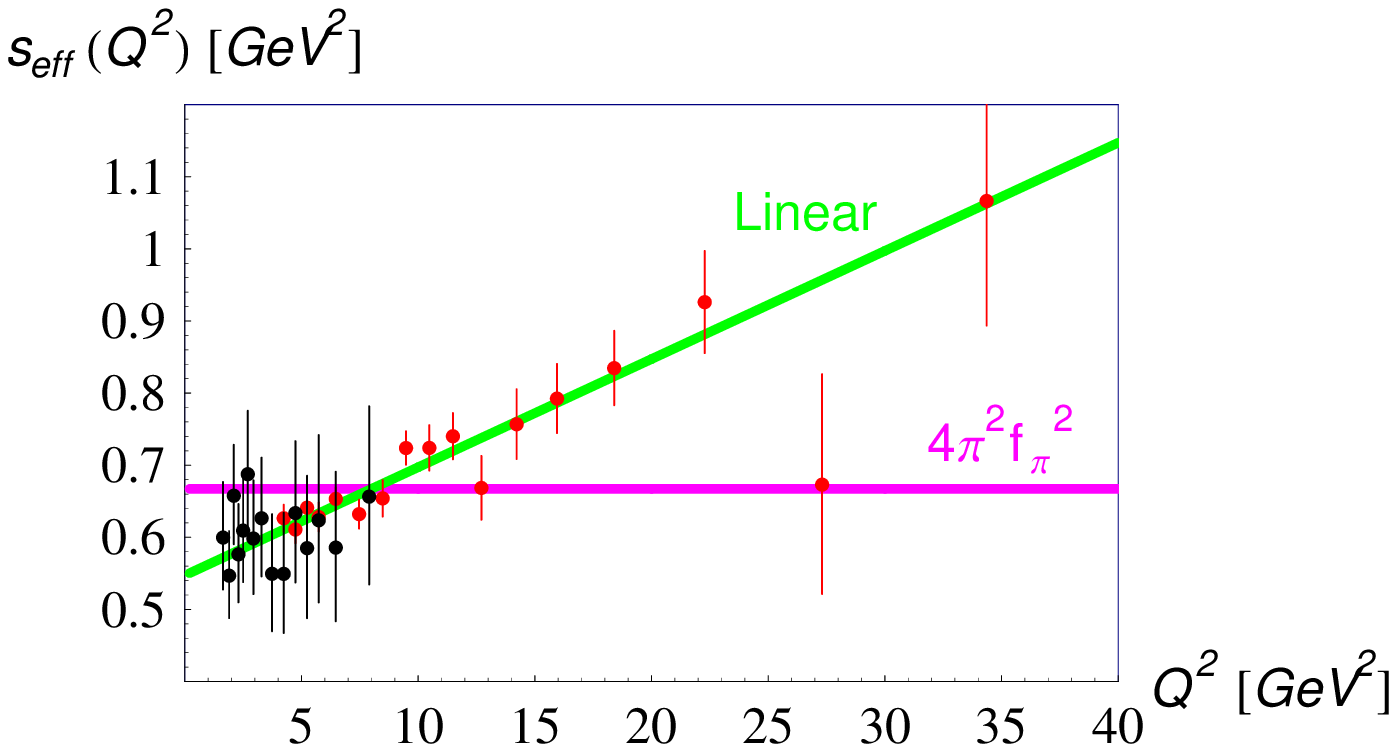}
\end{tabular}
\caption{\label{Plot:5}
The LD $\pi\gamma$ form factor vs. data from \cite{cello-cleo} (black) and  \cite{babar} (red) and the corresponding 
equivalent effective threshold.}
\end{center}
\end{figure}

Surprisingly, for the pion transition form factor, Fig.~\ref{Plot:5}, one observes a clear disagreement between 
the results from the LD model and the {\sc BaBar} data \cite{babar}. 
Moreover---in evident conflict with the $\eta$ and $\eta'$ results and the experience from quantum mechanics---the 
data implies that the violations of LD increase with $Q^2$ even in the region $Q^2\approx 40$ GeV$^2$! 
The effective threshold extracted from the {\sc BaBar} data is compatible with a linear growing function of $Q^2$ with no 
sign of approaching the LD limit. 

It is hard to find a convincing answer to the question why nonstrange components in $\eta$, $\eta'$, on the one hand, 
and in $\pi^0$, on the other hand, should behave so much differently?

%\newpage
%%%%%%%%%%%%%%%%%%%%%%%%%%%%%%%%%%%%%%%%%%%%%%%%%%%%%%%%%%%%%%%%%%%%%%%%%%%%%%%%%%%%%%%%%%%%%%%%%%%%%%%%%%%%%%%%%%
\section{Summary}
We presented the analysis of the pion elastic and the $\pi^0$, $\eta$, $\eta'$ transition form factors from the LD 
version of QCD sum rules. 
The main emphasis was laid on the attempt to probe the accuracy of this approximate method and the reliability of its predictions. 
%
%\noindent
Our main conclusions are as follows:

%\vspace{.4cm}
\begin{itemize}
\item
\noindent 
{\bf The elastic form factor}: 
Our quantum-mechanical analysis suggests that the LD model should work 
increasingly well in the region $Q^2\ge 4-8$ GeV$^2$, independently of the details of the confining interaction. 
For arbitrary confining interaction, the LD model gives very accurate results for $Q^2\ge 20-30$ GeV$^2$. 
The accurate data on the pion form factor at small momentum transfers indicate that the LD limit for the effective 
threshold, $s^{\rm LD}_{\rm eff}=4\pi^2 f_\pi^2$, may be reached already at relatively low values $Q^2=5-6$ GeV$^2$; 
thus, large deviations from the LD limit at $Q^2=20-50$~GeV$^2$ reported in some recent publications \cite{recent} 
appear to us rather unlikely. 
\item
\noindent 
{\bf The {\boldmath $P\to\gamma\gamma^*$} transition form factor}: 
We conclude from the quantum-mechanical analysis that %for a bound state of a usual hadron size $\simeq$ 1 fm, 
the LD model should work well in the region $Q^2\ge$ a few GeV$^2$. 
Indeed, for the $\eta\to\gamma\gamma^*$ and $\eta'\to\gamma\gamma^*$ form factors, the 
predictions from LD model in QCD work reasonably well. 
Surprisingly, for the $\pi\to \gamma\gamma^*$ form factor the present {\sc BaBar} data indicate an increasing violation 
of local duality, corresponding to 
a linearly rising effective threshold, even at $Q^2$ as large as 40 GeV$^2$. 
This puzzle has so far no compelling theoretical explanation. 
Our conclusion agrees with the findings of \cite{roberts,mikhailov,bt} obtained 
from other theoretical approaches.
\end{itemize}
\vspace{.5cm}

\noindent 
{\bf Acknowledgments.} 
We are grateful to S.~Mikhailov, O.~Nachtmann, O.~Teryaev, and particularly to B.~Stech for valuable discussions. 
D.~M.\ was supported by the Austrian Science Fund (FWF) under
Project No.~P22843 and is grateful to the Alexander von Humboldt-Stiftung and the Institute of 
Theoretical Physics of the Heidelberg University for financial support and hospitality during his stay in Heidelberg, 
where a part of this work was done.

\end{document}